\documentclass[runningheads]{llncs}
\usepackage[T1]{fontenc}
\usepackage{graphicx}
\usepackage{adjustbox}
\usepackage{multirow}
\usepackage{xcolor, colortbl}
\usepackage{booktabs}
\usepackage{subcaption}
\usepackage{amsmath}
\usepackage{amsfonts}
\usepackage{amssymb}
\usepackage{hyperref}

\definecolor{cgreen}{rgb}{0,0.7,0.6}
\definecolor{cred}{rgb}{0.968,0.145,0.521}
\newcommand{\rtrup}[1]{{\scriptsize\color{cred}$\blacktriangle$ #1}}
\newcommand{\rtr}[1]{{\scriptsize\color{cred}$\blacktriangledown$ #1}}
\newcommand{\gtrdown}[1]{{\scriptsize\color{cgreen}$\blacktriangledown$ #1}}
\newcommand{\gtr}[1]{{\scriptsize\color{cgreen}$\blacktriangle$ #1}}

\begin{document}
%
\title{Reversing the Abnormal: Pseudo-Healthy Generative Networks for Anomaly Detection}
%
\titlerunning{PHANES}
%
\author{Cosmin I. Bercea \inst{1,2} \and
Benedikt Wiestler \inst{3} \and
Daniel Rueckert \inst{1,3,4} \and
Julia A. Schnabel \inst{1,2, 5}
}
\authorrunning{CI. Bercea et al.}
\institute{Technical University of Munich, Germany \and
Helmholtz AI and Helmholtz Center Munich, Germany \and 
Klinikum Rechts der Isar, Munich, Germany \and 
Imperial College London, London, UK  \and 
King’s College London, London, UK 
}

%
%
\maketitle              
\begin{figure}[b!]
    \centering
    \includegraphics[width=0.643\textwidth]{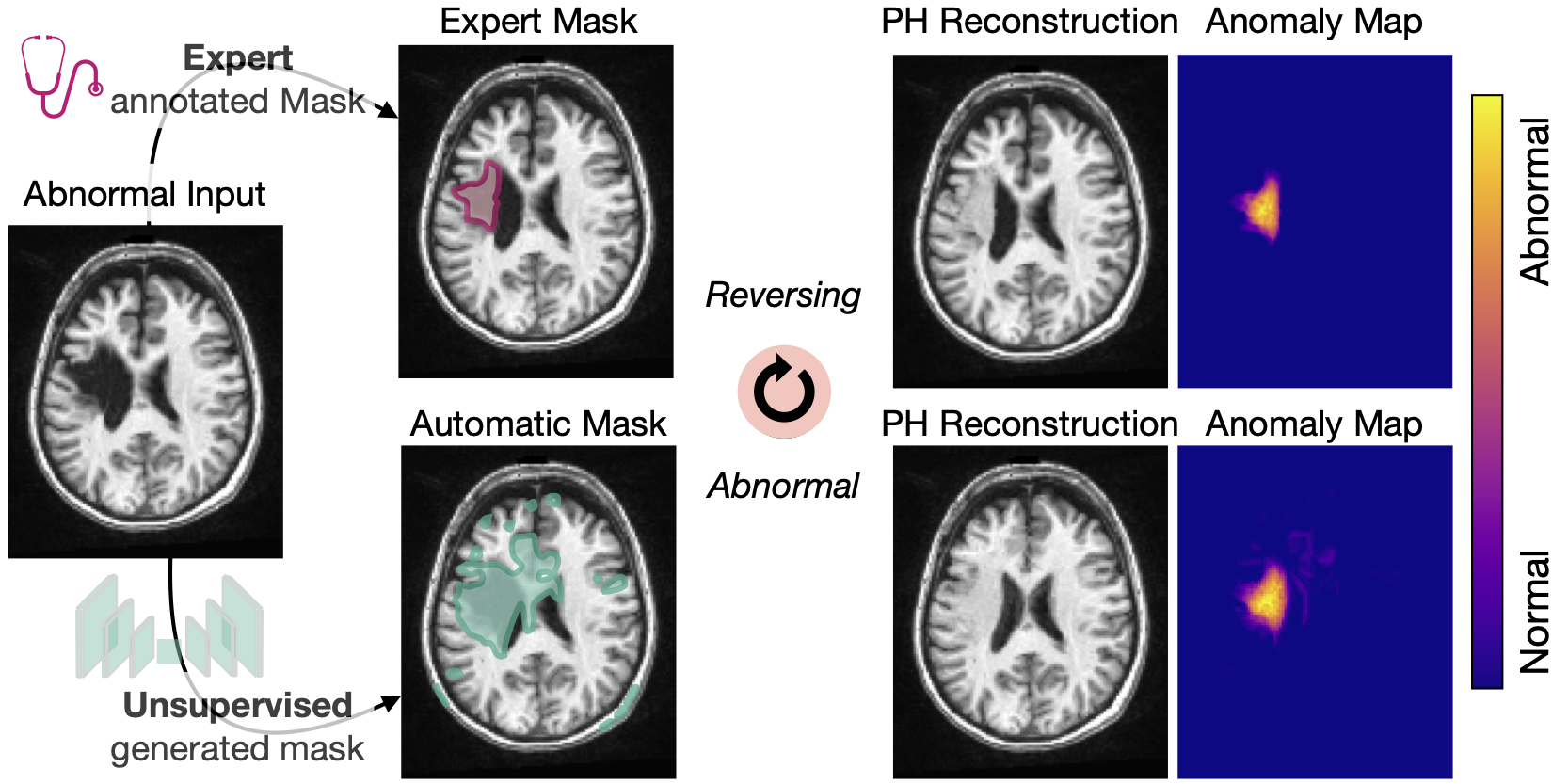} 
    \caption{Overview of \textit{PHANES}, see~\autoref{fig::phanes}. Our method can use both expert annotated- or unsupervised generated masks to reverse and segment anomalies.}
    \label{fig::phanes_teaser}
\end{figure}
\begin{abstract}
Early and accurate disease detection is crucial for patient management and successful treatment outcomes. However, the automatic identification of anomalies in medical images can be challenging. Conventional methods rely on large labeled datasets which are difficult to obtain. To overcome these limitations, we introduce a novel unsupervised approach, called \textit{PHANES} (Pseudo Healthy generative networks for ANomaly Segmentation). Our method has the capability of reversing anomalies, i.e., preserving healthy tissue and replacing anomalous regions with pseudo-healthy (PH) reconstructions. Unlike recent diffusion models, our method does not rely on a learned noise distribution nor does it introduce random alterations to the entire image. Instead, we use latent generative networks to create masks around possible anomalies, which are refined using inpainting generative networks. We demonstrate the effectiveness of \textit{PHANES} in detecting stroke lesions in T1w brain MRI datasets and show significant improvements over state-of-the-art (SOTA) methods. 
We believe that our proposed framework will open new avenues for interpretable, fast, and accurate anomaly segmentation with the potential to support various clinical-oriented downstream tasks. 
\end{abstract}
\section{Introduction}
The early detection of lesions in medical images is critical for the diagnosis and treatment of various conditions, including neurological disorders. Stroke is a leading cause of death and disability, where early detection and treatment can significantly improve patient outcomes. However, the quantification of lesion burden is challenging and can be time-consuming and subjective when performed manually by medical professionals~\cite{atlas2022}. 
While supervised learning methods~\cite{kamnitsas2016deepmedic,kamnitsas2017efficient} have proven to be effective in lesion segmentation, they rely heavily on large annotated datasets for training and tend to generalize poorly beyond the learned labels~\cite{ruff2021unifying}.
On the other hand, unsupervised methods focus on detecting patterns that significantly deviate from the norm by training only on normal data. 

One widely used category of unsupervised methods is latent restoration methods. They involve autoencoders (AEs) that learn low-dimensional representations of data and detect anomalies through inaccurate reconstructions of abnormal samples~\cite{pawlowski2018unsupervised}. However, developing compact and comprehensive representations of the healthy distribution is challenging~\cite{bercea2022ra}, as recent studies suggest AEs perform better reconstructions on out-of-distribution (OOD) samples than on training samples~\cite{schirrmeister2020understanding}. 
Various techniques have been introduced to enhance representation learning, including discretizing the latent space~\cite{mao2020abnormality}, disentangling compounding factors~\cite{bercea2022fedano}, and variational autoencoders (VAEs) that introduce a prior into the latent distribution~\cite{you2019unsupervised,zimmerer2019unsupervised}. However, methods that can enforce the reconstruction of healthy generally tend to produce blurry reconstructions.

In contrast, generative adversarial networks (GANs)\cite{goodfellow2014generative,perera2019ocgan,schlegl2019fanogan} are capable of producing high-resolution images. New adversarial AEs combine VAEs' latent representations with GANs' generative abilities, achieving SOTA results in image generation and outlier detection\cite{chen2018unsupervised,daniel2021soft,bercea2022ra}. 
Nevertheless, latent methods still face difficulties in accurately reconstructing data from their low-dimensional representations, causing false positive detections on healthy tissues.

Several techniques have been proposed that make use of the inherent spatial information in the data rather than relying on constrained latent representations~\cite{kascenas2022denoising,Wyatt_2022_CVPR,zimmerer2018context}. 
These methods are often trained on a pretext task, such as recovering masked input content~\cite{zimmerer2018context}. De-noising AEs~\cite{kascenas2022denoising} are trained to eliminate synthetic noise patterns, utilizing skip connections to preserve the spatial information and achieve SOTA brain tumor segmentation. However, they heavily rely on a learned noise model and may miss anomalies that deviate from the noise distribution~\cite{bercea2022ra}. More recently, diffusion models~\cite{ho2020denoising} apply a more complex de-noising process to detect anomalies~\cite{Wyatt_2022_CVPR}. However, the choice and granularity of the applied noise is crucial for breaking the structure of anomalies~\cite{Wyatt_2022_CVPR}. Adapting the noise distribution to the diversity and heterogeneity of pathology is inherently difficult, and even if achieved, the noising process disrupts the structure of both healthy and anomalous regions throughout the entire image.

In related computer vision areas, such as industrial inspection~\cite{mvtec1}, the top-performing methods do not focus on reversing anomalies, but rather on detecting them by using large nominal banks~\cite{defard2021padim,roth2022towards}, or pre-trained features from large natural imaging datasets like ImageNet~\cite{bergmann2020uninformed,salehi2021multiresolution}. Salehi et al.~\cite{salehi2021multiresolution} have employed multi-scale knowledge distillation to detect anomalies in industrial and medical imaging. However, the application of these networks in medical anomaly segmentation, particularly in brain MRI, is limited by various challenges specific to the medical imaging domain. They include the variability and complexity of normal data, subtlety of anomalies, limited size of datasets, and domain shifts.

This work aims to combine the advantages of constrained latent restoration for understanding healthy data distribution with generative in-painting networks. Unlike previous methods, our approach does not rely on a learned noise model, but instead creates masks of probable anomalies using latent restoration. These guide generative in-painting networks to reverse anomalies, i.e., preserve healthy tissues and produce pseudo-healthy in-painting in anomalous regions. 
We believe that our proposed method will open new avenues for interpretable, fast, and accurate anomaly segmentation and support various clinical-oriented downstream tasks, such as investigating progression of disease, patient stratification and treatment planning. In summary our main contributions are: 
\begin{itemize}
    \item[$\bullet$] We investigate and measure the ability of SOTA methods to reverse synthetic anomalies on real brain T1w MRI data.
    \item[$\bullet$] We propose a novel unsupervised segmentation framework, that we call \textit{PHANES}, that is able to preserve healthy regions and utilize them to generate pseudo-healthy reconstructions on anomalous regions.
    \item[$\bullet$] We demonstrate a significant advancement in the challenging task of unsupervised ischemic stroke lesion segmentation.
\end{itemize}
\section{Background}
\textbf{Latent restoration methods} use neural networks to estimate the parameters $\theta,\phi$ of an encoder $E_{\theta}$ and a decoder $D_{\phi}$. The aim is to restore the input from its lower-dimensional latent representation with minimal loss. The standard objective is to minimize the residual, e.g., using mean squared error (MSE) loss: $\min_{\theta,\phi}\sum_{i=1}^N \| x_i - D_\phi(E_\theta(x_i))\|^2$. In the context of variational inference~\cite{kingma2013auto}, the goal is to optimize the parameters $\theta$ of a latent variable model $p_{\theta}$(x) by maximizing the log-likelihood of the observed samples $x$: $\log p_{\theta}(x)$. The term is intractable, but the true posterior $p_{\theta}(z|x)$ can be approximated by $q_{\phi}(z|x)$:
 \begin{equation}
    \label{eq::elbo}
    \log p_\theta(x)  \geq \mathbb{E}_{q(z|x)} [\log p_\theta(x|z)] - KL[q_\phi(z|x) || p(z)] = ELBO(x).
\end{equation}
KL is the Kullback-Leibler divergence;  $q_\phi(z|x)$ and $p_\theta(x|z)$ are usually known as the encoder $E_\phi$ and decoder $D_\theta$; the prior $p(z)$ is usually the normal distribution $\mathcal{N}(\mu_0, \sigma_0)$; and the ELBO denotes the Evidence Lower Bound. In unsupervised anomaly detection, the networks are trained only on normal samples $x \in \mathcal{X} \subset \mathbb{R}^N$. Given an anomalous input $\overline{x} \notin \mathcal{X}$, it is assumed that the reconstruction $x_{ph}=(D_\phi(E_\theta(\overline{x}))) \in \mathcal{X}$ represents its pseudo-healthy version. The aim of the pseudo-healthy reconstructions is to accurately reverse abnormalities present in the input images. This is achieved by preserving the healthy regions while generating healthy-like tissues in anomalous regions. Thus, anomalies can ideally be directly localized by computing the difference between the anomalous input and the pseudo-healthy reconstructions: $s(\overline{x}) = |\overline{x}-x_{ph}|$. 

\section{Method}
\begin{figure}[tb]
    \centering
    \includegraphics[width=\textwidth]{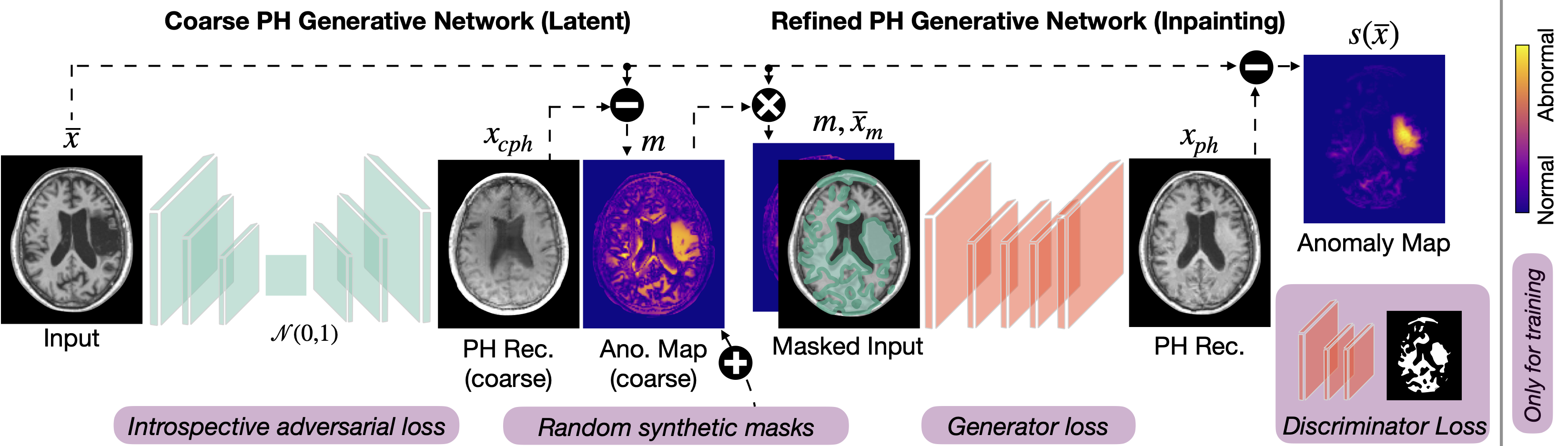}
    \caption{\textit{PHANES} overview. Our framework offers modularity, enabling the choice of preferred generative networks, such as adversarial or diffusion-based models. First, we use latent generative networks to learn the healthy data distribution and provide approximate pseudo-healthy reconstructions $x_{cph}$. Anomaly maps $m$ obtained from this step are then used to mask out possible anomalous regions in the input. The remaining healthy tissues are used to condition the refined generative networks, which complete the image and replace anomalous regions with pseudo-healthy tissues. This results in accurate PH reconstructions $x_{ph}$, which enables the precise localization of diseases, as shown on the right.}
    \label{fig::phanes}
\end{figure}

~\autoref{fig::phanes} shows an overview of our proposed method. We introduce an innovative approach by utilizing masks produced by latent generative networks to condition generative inpainting networks only on healthy tissues. Our framework is modular, which allows for the flexibility of choosing a preferred generative network, such as adversarial, or diffusion-based models for predicting the pseudo-healthy reconstructions. In the following we describe each component in detail.

\textbf{Latent generative network.} The first step involves generating masks to cover potential anomalous regions in the input image. The goal of this step is to achieve unbiased detection of various pathologies and minimize false positives. It is therefore important to use a method that is restricted to in-distribution samples, particularly healthy samples, while also accurately reconstructing inputs. Here, we have adopted our previous work~\cite{bercea2022ra} that augments a soft introspective variational auto-encoder with a reversed embedding similarity loss with the aim to enforcing more accurate pseudo-healthy reconstructions. The training process encourages the encoder to distinguish between real and generated samples by minimizing the Kullback-Leibler (KL) divergence of the latent distribution of real samples and the prior, and maximizing the KL divergence of generated samples. On the other hand, the decoder is trained to deceive the encoder by reconstructing real data samples using the standard ELBO and minimizing the KL divergence of generated samples compressed by the encoder:  
\begin{flalign}
    \label{eq::ra}
    & \mathcal{L}_{E_{\phi}}(x,z) = ELBO(x) - \frac{1}{\alpha}(exp(\alpha ELBO(D_\theta(z)) + \lambda \mathcal{L}_{Reversed}(x),\\
    & \mathcal{L}_{Reversed}(x) = \sum_{l=0}^L (1 - \mathcal{L}_{Sim}(E_\phi^l(x), E_\phi^l(x_{cph})) + \frac{1}{2} MSE(E_\phi^l(x), E_\phi^l(x_{cph})),\nonumber\\ 
    & \mathcal{L}_{D_{\theta}}(x,z) = ELBO(x) + \gamma ELBO(D_\theta(z)), \nonumber
\end{flalign}
where $E_\phi^l$ is the $l$-th embedding of the $L$ encoder layers, $x_{cph}=D_\theta(E_\phi(x))$, and $\mathcal{L}_{Sim}$ is the cosine similarity.

\textbf{Mask generation.} Simple residual errors have a strong dependence on the underlying intensities~\cite{meissen2022pitfalls}. As it is important to assign higher values to (subtle) pathological structures, we compute anomaly masks as proposed in~\cite{bercea2022ra} by applying adaptive histogram equalization (eq), normalizing with the $95th$ percentile, and augmenting the errors with perceptual differences for robustness:
\begin{equation}
    m(\overline{x}) = norm_{95}(|(eq(x_{cph}) - eq(\overline{x})|) * {\cal S}_{lpips}(eq(x_{cph}), eq(\overline{x})),
\end{equation}
with ${\cal S}_{lpips}$ being the learned perceptual image patch similarity~\cite{zhang2018unreasonable}. Finally, we binarize the masks using the 99th percentile value on the healthy validation set. 

\textbf{Inpainting generative network}. The objective of the refined PH generative network is to complete the masked image by utilizing the remaining healthy tissues to generate a full PH version of the input. Considering computational efficiency, we have employed the recent in-painting AOT-GAN~\cite{zeng2022aggregated}. The method uses a generator ($G$) and discriminator neural network to optimize losses based on residual and perceptual differences, resulting in accurate and visually precise inpainted images. Additionally, the discriminator predicts the input mask from the inpainted image to improve the synthesis of fine textures.

\textbf{Anomaly Maps.} The PH reconstruction is computed as follows: $x_{ph} = \overline{x} \odot (1-m) + G(\overline{x}\odot (1-m), m) \odot m$, with $\odot$ being the pixel-wise multiplication. We compute the final anomaly maps based on residual and perceptual differences:
\begin{equation}
    \label{eq::ano}
    s(\overline{x}) = |x_{ph} - \overline{x}| * {\cal S}_{lpips}(x_{ph}, \overline{x})
\end{equation}
\section{Experiments}
\textbf{Datasets.} We trained our model using two publicly available brain T1w MRI datasets, including FastMRI+ (131 train, 15 val, 30 test) and IXI (581 train samples), to capture the healthy distribution. Performance evaluation was done on a large stroke T1-weighted MRI dataset, ATLAS v2.0~\cite{atlas2022}, containing 655 images with manually segmented lesion masks for training and 355 test images with hidden lesion masks. We evaluated the model using the 655 training images with public annotations. The mid axial slices were normalized to the $98^{th}$ percentile, padded, and resized to $128\times128$ resolution. During training, we performed random rotations up to 10 degrees, translations up to 0.1, scaling from 0.9 to 1.1, and horizontal flips with a 0.5 probability. We trained for 1500 epochs, with a batch size of 8, lr of $5e^{-5}$, and early stopping (see Appendix for details). 
\subsection{Reversing Synthetic Anomalies\label{sec::synth}}
\begin{table}[t!]
    \centering
    \setlength{\tabcolsep}{6pt}
    \caption{\textbf{Reversing synthetic anomalies.} We evaluate the pseudo-healthy (PH) reconstruction on healthy and anomalous regions using the learned perceptual image patch similarity (LPIPS)~\cite{zhang2018unreasonable} and the anomaly segmentation performance. \textit{PHANES}$^{GT}$ represents an upper bound and uses the ground truth anomalies to mask the image for inpainting.~{\color{cgreen}x$\%$} shows improvement over best baseline (RA) and~{\color{cred}x$\%$} shows the decrease in performance compared to \textit{PHANES}.\label{tab::synth_phr}}
    \begin{adjustbox}{width=0.85\linewidth,center} 
        \begin{tabular}{l | c c || c c}
            \toprule	    
            \multirow{2}{*}{Method} & \multicolumn{2}{c||}{PH Reconstruction (LPIPS)} & \multicolumn{2}{c}{Anomaly Segmentation}\\
            & Healthy $\downarrow$& Anomaly $\downarrow$& AUPRC $\uparrow$ & $\lceil DICE \rceil$ $\uparrow$
            \\\midrule
            \rowcolor{gray!10} PHANES$^{GT}$ (ours) &  {\boldmath $0.09$}~{\scriptsize\color{cgreen} N/A}  & {\boldmath$0.94$}~\gtrdown{94\%} & {\boldmath$100$}~\gtr{37\%} & {\boldmath$100$}~\gtr{46\%}\\
            \rowcolor{gray!10} PHANES (ours) & {\boldmath$2.25$}~\gtrdown{77\%} &  {\boldmath$8.10$}~\gtrdown{47\%}&{\boldmath$77.93$}~\gtr{7\%}  &{\boldmath$75.47$}~\gtr{10\%}  \\ \hline
            RA~\cite{bercea2022ra} & $9.74$~\rtrup{333\%} & $15.27$~\rtrup{89\%} & $73.01$~\rtr{6\%} & $68.52$~\rtr{9\%} \\
            SI-VAE~\cite{daniel2021soft} & $13.16$~\rtrup{485\%}  & $19.01$~\rtrup{135\%} & $17.91$~\rtr{77\%} & $31.30$~\rtr{59\%} \\
     	   AnoDDPM~\cite{Wyatt_2022_CVPR} & $6.64$~\rtrup{195\%} & $19.46$~\rtrup{140\%}  & $14.85$~\rtr{81\%} & $19.89$~\rtr{74\%}\\ 
     	     DAE~\cite{kascenas2022denoising}  &  $3.94$~\rtrup{75\%}  & $20.05$~\rtrup{148\%} & $35.73$~\rtr{54\%} & $37.76$~\rtr{50\%}  \\
    	    VAE~\cite{zimmerer2019unsupervised} & $33.22$~\rtrup{1376\%} & $44.00$~\rtrup{443\%} & $22.86$~\rtr{71\%} & $28.46$~\rtr{62\%} \\
     	    \bottomrule
        \end{tabular}
    \end{adjustbox}
\end{table}
\begin{figure}[tb]
    \centering
    \includegraphics[width=0.8\textwidth]{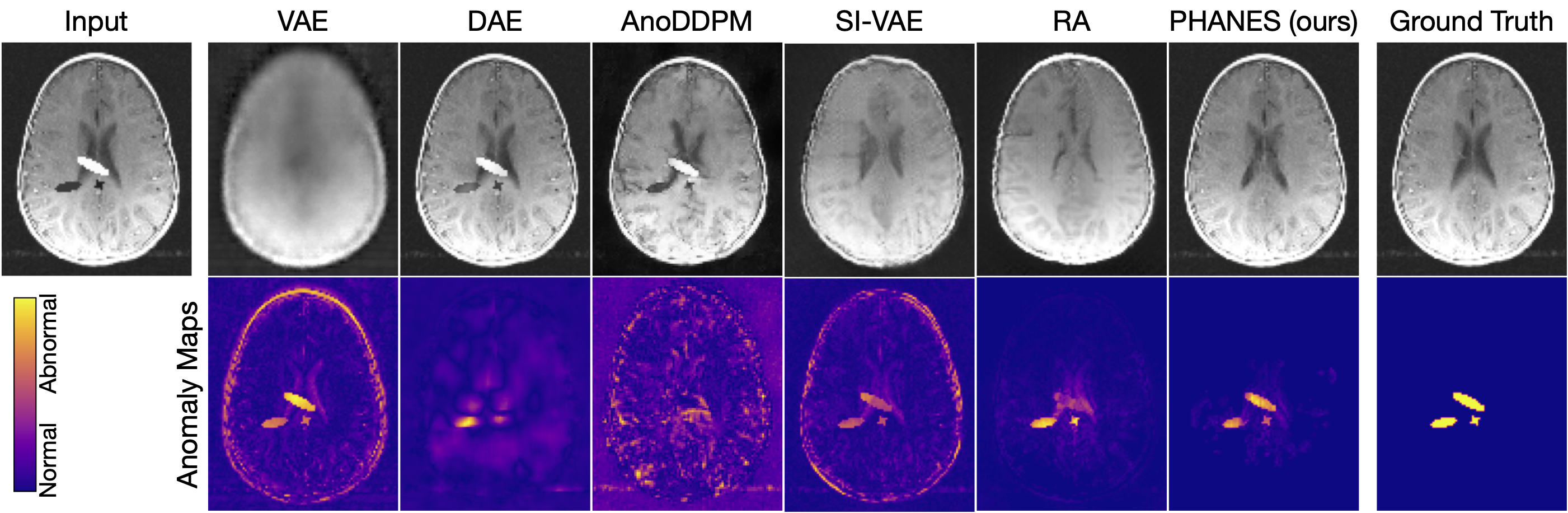}
    \caption{\textbf{Reversing synthetic anomalies.} \textit{PHANES} successfully removes synthetic anomalies and produces the most accurate pseudo-healthy reconstructions. }
    \label{fig::phr}
\end{figure}
In this section, we test whether reconstruction-based methods can generate pseudo-healthy images and reverse synthetic anomalies. Results are evaluated in~\autoref{tab::synth_phr} and~\autoref{fig::phr} using 30 test images and synthetic masks as reference. VAEs produce blurry results that lead to poor reconstruction of both healthy and anomalous regions (LPIPS) and thus poor segmentation performance. While DAEs preserve the healthy tissues well with an LPIPS of 3.94, they do not generate pseudo-healthy reconstructions in anomalous regions (LPIPS $\approx20$). However, they change the intensity of some structures, e.g., hypo-intensities, allowing for improved detection accuracy (see AUPRC and Dice). Simplex noise in~\cite{Wyatt_2022_CVPR} is designed to detect large hypo-intense lesions, leaving small anomalies undetected by AnoDDPM. SI-VAEs and RA produce pseudo healthy versions of the abnormal inputs, with the latter achieving the best results among the baselines. Our proposed method, \textit{PHANES}, successfully reverses the synthetic anomalies, with its reconstructions being the most similar to ground truth healthy samples, as can be seen in~\autoref{fig::phr}. It achieved an improvement of 77\% and 47\% in generating pseudo-healthy samples in healthy and anomalous regions, respectfully. This enables the precised localization of anomalies (see bottom row in~\autoref{fig::phr}). 
\subsection{Ischemic Stroke Lesion Segmentation on T1w Brain MRI}
\begin{table}[t]
    \caption{\textbf{Ischemic stroke lesion segmentation on real T1w brain MRIs.}\\
    ~\gtr{x$\%$} shows improvement over AnoDDPM, and~\rtr{x$\%$} shows the decrease in performance compared to \textit{PHANES}. $^*$ marks statistical significance ($p<0.05$).\label{tab::benchmark_anomaly_detection}}
    \centering
    \setlength{\tabcolsep}{10pt}
        \begin{adjustbox}{width=0.75\linewidth,center} 
            \centering
            \begin{tabular}{l | c c }
                \toprule	    
                \multirow{1}{*}{Method}  &\multicolumn{1}{c}{AUPRC $\uparrow$} &\multicolumn{1}{c}{$\lceil DICE \rceil$ $\uparrow$}\\\midrule
                \rowcolor{gray!10} PHANES (ours) & {\boldmath $19.96 \pm 2.3^*$}~\gtr{22$\%$} & {\boldmath $32.17\pm 2.0^*$}~\gtr{16$\%$}  \\\hline
                AnoDDPM~\cite{Wyatt_2022_CVPR}  & $16.33\pm1.7$~\rtr{18$\%$} & $27.64\pm1.4$~\rtr{14$\%$}\\
                RA~\cite{bercea2022ra}  & $12.30\pm1.0$~\rtr{38$\%$} & $22.20\pm1.5$~\rtr{31$\%$} \\
                PatchCore~\cite{roth2022towards} & $12.24\pm 0.7$~\rtr{39$\%$} & $24.79\pm1.2$~\rtr{23$\%$} \\
          	    DAE~\cite{kascenas2022denoising}   & $9.22\pm1.3$~\rtr{54$\%$} & $15.62\pm2.1$~\rtr{53$\%$} \\
	            SI-VAE~\cite{daniel2021soft} & $6.86\pm0.6$~\rtr{66$\%$} & $13.57\pm0.9$~\rtr{58$\%$} \\
          	    MKD~\cite{salehi2021multiresolution} & $2.93\pm0.3$~\rtr{85$\%$} & $5.91\pm0.6$~\rtr{82$\%$}\\
        	    VAE~\cite{zimmerer2019unsupervised} & $2.76\pm0.2$~\rtr{86$\%$} & $5.96\pm0.3$~\rtr{81$\%$}\\
         	    \bottomrule
            \end{tabular}
        \end{adjustbox}
\end{table}

\begin{figure}[tb]
    \centering
    \includegraphics[width=\textwidth]{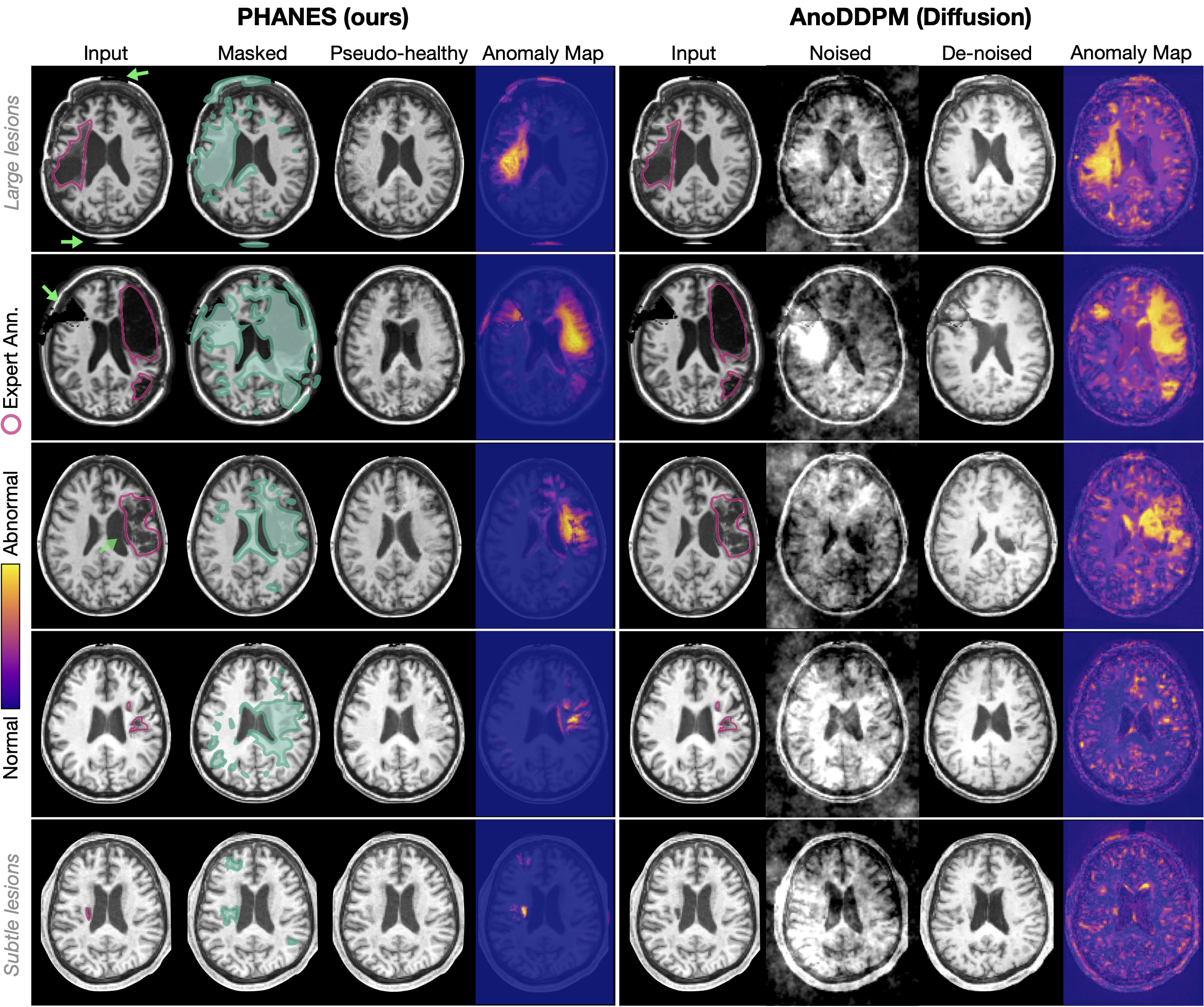}
    \caption{\textbf{Stroke lesion segmentation.} We show input images with expert annotations in red along with masked images generated by the latent generative networks in~\autoref{fig::phanes}, pseudo-healthy reconstructions, and anomaly maps. On the right, we show the performance of diffusion models on the same inputs. Different rows show cases ranging from large anomalies at the top to more subtle ones at the bottom. Green arrows mark unlabeled anomalies. \textit{PHANES} successfully reverses the anomalies and accurately localizes even very subtle anomalies.}
    \label{fig::quali}
\end{figure}
In this section, we evaluate the performance of our approach in segmenting stroke lesions and show the results in~\autoref{tab::benchmark_anomaly_detection} and ~\autoref{fig::quali}. For completeness, we compare our approach to teacher-student methods that use multi-scale knowledge distillation (MKD) for anomaly segmentation. The unsupervised detection of (subtle) stroke lesions is challenging. The lack of healthy data from the same scanner and the limited size of the healthy datasets limit the successful application of such methods, with a maximum achievable Dice score of just under $6\%$. On the other hand, PatchCore, which is currently the SOTA method in industrial anomaly detection, has demonstrated comparable performance to the top-performing baselines. VAEs yield many false positive detections due to the blurry reconstructions and achieve poor localization results. DAEs can identify anomalies that resemble the learned noise distribution and improve segmentation results (AUPRC of $9.22$), despite not producing pseudo-healthy reconstructions of abnormal samples (see~\autoref{sec::synth}). The best performing latent restoration method is RA, achieving a remarkable $79\%$ improvement over SI-VAE. Unlike experiments in~\autoref{sec::synth}, the Simplex noise aligns more closely with the hypo-intense pathology distribution of stroke in T1w brain MRI. As a result, AnoDDPM achieves the highest detection accuracy among the baselines. Compared to AnoDDPM, \textit{PHANES} increases the detection results by $22\%$ AUPRC.~\autoref{fig::quali} shows a visual comparison between the two approaches. Diffusion models tend to be more susceptible to domain shifts (first three rows) and yield more false positives. In contrast, \textit{PHANES} demonstrates more precise localization, especially for subtle anomalies (bottom rows). Generally, unsupervised methods tend to have lower Dice scores partly due to unlabeled artefacts in the dataset. These include non-pathological (rows 1,2) as well as other pathological effects, such as changes in ventricle structure (rows 3,4). \textit{PHANES} correctly identifies these as anomalous, but their lack of annotation limits numerical evaluations.
\section{Discussion}
This paper presents a novel unsupervised anomaly segmentation framework, called \textit{PHANES}. It possesses the ability to reverse anomalies in medical images by preserving healthy tissues and substituting anomalous regions with pseudo-healthy reconstructions. By generating pseudo-healthy versions of images containing anomalies, \textit{PHANES} can be a useful tool in supporting clinical studies. In this work, we demonstrated exceptional performance in reversing synthetic anomalies and segmenting stroke lesions on brain T1w MRIs. We believe that deliberate masking of (possible) abnormal regions will pave new ways for novel anomaly segmentation methods and empower further clinical applications.
%
%
%
\bibliographystyle{splncs04}
\bibliography{main}

\begin{thebibliography}{10}
\providecommand{\url}[1]{\texttt{#1}}
\providecommand{\urlprefix}{URL }
\providecommand{\doi}[1]{https://doi.org/#1}

\bibitem{bercea2022ra}
Bercea, C.I., Wiestler, B., Rueckert, D., A, S.J.: Generalizing unsupervised
  anomaly detection: Towards unbiased pathology screening. International
  Conference on Medical Imaging with Deep Learning  (2023)

\bibitem{bercea2022fedano}
Bercea, C.I., Wiestler, B., Rueckert, D., Albarqouni, S.: Federated
  disentangled representation learning for unsupervised brain anomaly
  detection. Nature Machine Intelligence  \textbf{4}(8),  685--695 (2022)

\bibitem{mvtec1}
Bergmann, P., Fauser, M., Sattlegger, D., Steger, C.: {MVT}ec {AD} — a
  comprehensive real-world dataset for unsupervised anomaly detection. In:
  Proceedings of the IEEE/CVF Conference on Computer Vision and Pattern
  Recognition. pp. 9584--9592 (2019)

\bibitem{bergmann2020uninformed}
Bergmann, P., Fauser, M., Sattlegger, D., Steger, C.: Uninformed students:
  Student-teacher anomaly detection with discriminative latent embeddings. In:
  Proceedings of the IEEE/CVF Conference on Computer Vision and Pattern
  Recognition. pp. 4183--4192 (2020)

\bibitem{chen2018unsupervised}
Chen, X., Konukoglu, E.: Unsupervised detection of lesions in brain {MRI} using
  constrained adversarial auto-encoders. In: International Conference on
  Medical Imaging with Deep Learning (2018)

\bibitem{daniel2021soft}
Daniel, T., Tamar, A.: Soft-{I}ntro{VAE}: Analyzing and improving the
  introspective variational autoencoder. In: Proceedings of the IEEE/CVF
  Conference on Computer Vision and Pattern Recognition. pp. 4391--4400 (2021)

\bibitem{defard2021padim}
Defard, T., Setkov, A., Loesch, A., Audigier, R.: Padim: a patch distribution
  modeling framework for anomaly detection and localization. In: Pattern
  Recognition. ICPR International Workshops and Challenges. pp. 475--489.
  Springer (2021)

\bibitem{goodfellow2014generative}
Goodfellow, I., Pouget-Abadie, J., Mirza, M., Xu, B., Warde-Farley, D., Ozair,
  S., Courville, A., Bengio, Y.: Generative adversarial nets. Advances in
  Neural Information Processing Systems  \textbf{27} (2014)

\bibitem{ho2020denoising}
Ho, J., Jain, A., Abbeel, P.: Denoising diffusion probabilistic models.
  Advances in Neural Information Processing Systems  \textbf{33},  6840--6851
  (2020)

\bibitem{kamnitsas2016deepmedic}
Kamnitsas, K., Ferrante, E., Parisot, S., Ledig, C., Nori, A.V., Criminisi, A.,
  Rueckert, D., Glocker, B.: Deep{M}edic for brain tumor segmentation. In:
  Medical Image Computing and Computer Assisted Intervention BrainLes Workshop.
  pp. 138--149 (2016)

\bibitem{kamnitsas2017efficient}
Kamnitsas, K., Ledig, C., Newcombe, V.F., Simpson, J.P., Kane, A.D., Menon,
  D.K., Rueckert, D., Glocker, B.: Efficient multi-scale {3D} {CNN} with fully
  connected {CRF} for accurate brain lesion segmentation. Medical Image
  Analysis  \textbf{36},  61--78 (2017)

\bibitem{kascenas2022denoising}
Kascenas, A., Pugeault, N., O'Neil, A.Q.: Denoising autoencoders for
  unsupervised anomaly detection in brain {MRI}. In: International Conference
  on Medical Imaging with Deep Learning (2022)

\bibitem{kingma2013auto}
Kingma, D.P., Welling, M.: Auto-encoding variational {B}ayes. arXiv preprint
  arXiv:1312.6114  (2013)

\bibitem{atlas2022}
Liew, S.L., Lo, B.P., ., Miarnda R.~Donnelly, e.a.: A large, curated,
  open-source stroke neuroimaging dataset to improve lesion segmentation
  algorithms. Scientific Data  \textbf{9} (2022)

\bibitem{mao2020abnormality}
Mao, Y., Xue, F.F., Wang, R., Zhang, J., Zheng, W.S., Liu, H.: Abnormality
  detection in chest {X}-ray images using uncertainty prediction autoencoders.
  In: Medical Image Computing and Computer Assisted Intervention. pp. 529--538.
  Springer (2020)

\bibitem{meissen2022pitfalls}
Meissen, F., Wiestler, B., Kaissis, G., Rueckert, D.: On the pitfalls of using
  the residual error as anomaly score. arXiv preprint arXiv:2202.03826  (2022)

\bibitem{pawlowski2018unsupervised}
Pawlowski, N., Lee, M.C., Rajchl, M., McDonagh, S., Ferrante, E., Kamnitsas,
  K., Cooke, S., Stevenson, S., Khetani, A., Newman, T., et~al.: Unsupervised
  lesion detection in brain {CT} using {B}ayesian convolutional autoencoders.
  International Conference on Medical Imaging with Deep Learning  (2018)

\bibitem{perera2019ocgan}
Perera, P., Nallapati, R., Xiang, B.: Ocgan: One-class novelty detection using
  gans with constrained latent representations. In: Proceedings of the IEEE/CVF
  Conference on Computer Vision and Pattern Recognition. pp. 2898--2906 (2019)

\bibitem{roth2022towards}
Roth, K., Pemula, L., Zepeda, J., Sch{\"o}lkopf, B., Brox, T., Gehler, P.:
  Towards total recall in industrial anomaly detection. In: Proceedings of the
  IEEE/CVF Conference on Computer Vision and Pattern Recognition. pp.
  14318--14328 (2022)

\bibitem{ruff2021unifying}
Ruff, L., Kauffmann, J.R., Vandermeulen, R.A., Montavon, G., Samek, W., Kloft,
  M., Dietterich, T.G., M{\"u}ller, K.R.: A unifying review of deep and shallow
  anomaly detection. Proc. IEEE  (2021)

\bibitem{salehi2021multiresolution}
Salehi, M., Sadjadi, N., Baselizadeh, S., Rohban, M.H., Rabiee, H.R.:
  Multiresolution knowledge distillation for anomaly detection. In: Proceedings
  of the IEEE/CVF Conference on Computer Vision and Pattern Recognition. pp.
  14902--14912 (2021)

\bibitem{schirrmeister2020understanding}
Schirrmeister, R., Zhou, Y., Ball, T., Zhang, D.: Understanding anomaly
  detection with deep invertible networks through hierarchies of distributions
  and features. Advances in Neural Information Processing Systems  \textbf{33},
   21038--21049 (2020)

\bibitem{schlegl2019fanogan}
Schlegl, T., Seeböck, P., Waldstein, S.M., Langs, G., Schmidt-Erfurth, U.:
  f-{A}no{GAN}: Fast unsupervised anomaly detection with generative adversarial
  networks. Medical Image Analysis  \textbf{54},  30--44 (2019)

\bibitem{Wyatt_2022_CVPR}
Wyatt, J., Leach, A., Schmon, S.M., Willcocks, C.G.: Anoddpm: Anomaly detection
  with denoising diffusion probabilistic models using simplex noise. In:
  Proceedings of the IEEE/CVF Conference on Computer Vision and Pattern
  Recognition Workshops. pp. 650--656 (June 2022)

\bibitem{you2019unsupervised}
You, S., Tezcan, K.C., Chen, X., Konukoglu, E.: Unsupervised lesion detection
  via image restoration with a normative prior. In: International Conference on
  Medical Imaging with Deep Learning. pp. 540--556. PMLR (2019)

\bibitem{zeng2022aggregated}
Zeng, Y., Fu, J., Chao, H., Guo, B.: Aggregated contextual transformations for
  high-resolution image inpainting. IEEE Transactions on Visualization and
  Computer Graphics  (2022)

\bibitem{zhang2018unreasonable}
Zhang, R., Isola, P., Efros, A.A., Shechtman, E., Wang, O.: The unreasonable
  effectiveness of deep features as a perceptual metric. In: Proceedings of the
  IEEE/CVF Conference on Computer Vision and Pattern Recognition. pp. 586--595
  (2018)

\bibitem{zimmerer2019unsupervised}
Zimmerer, D., Isensee, F., Petersen, J., Kohl, S., Maier-Hein, K.: Unsupervised
  anomaly localization using variational auto-encoders. In: Medical Image
  Computing and Computer Assisted Intervention. pp. 289--297. Springer (2019)

\bibitem{zimmerer2018context}
Zimmerer, D., Kohl, S.A., Petersen, J., Isensee, F., Maier-Hein, K.H.:
  Context-encoding variational autoencoder for unsupervised anomaly detection.
  arXiv preprint arXiv:1812.05941  (2018)

\end{thebibliography}
\section{Appendix}
\begin{figure}[htbp]
    \begin{minipage}{\textwidth}
    \centering
        \includegraphics[width=0.9\textwidth]{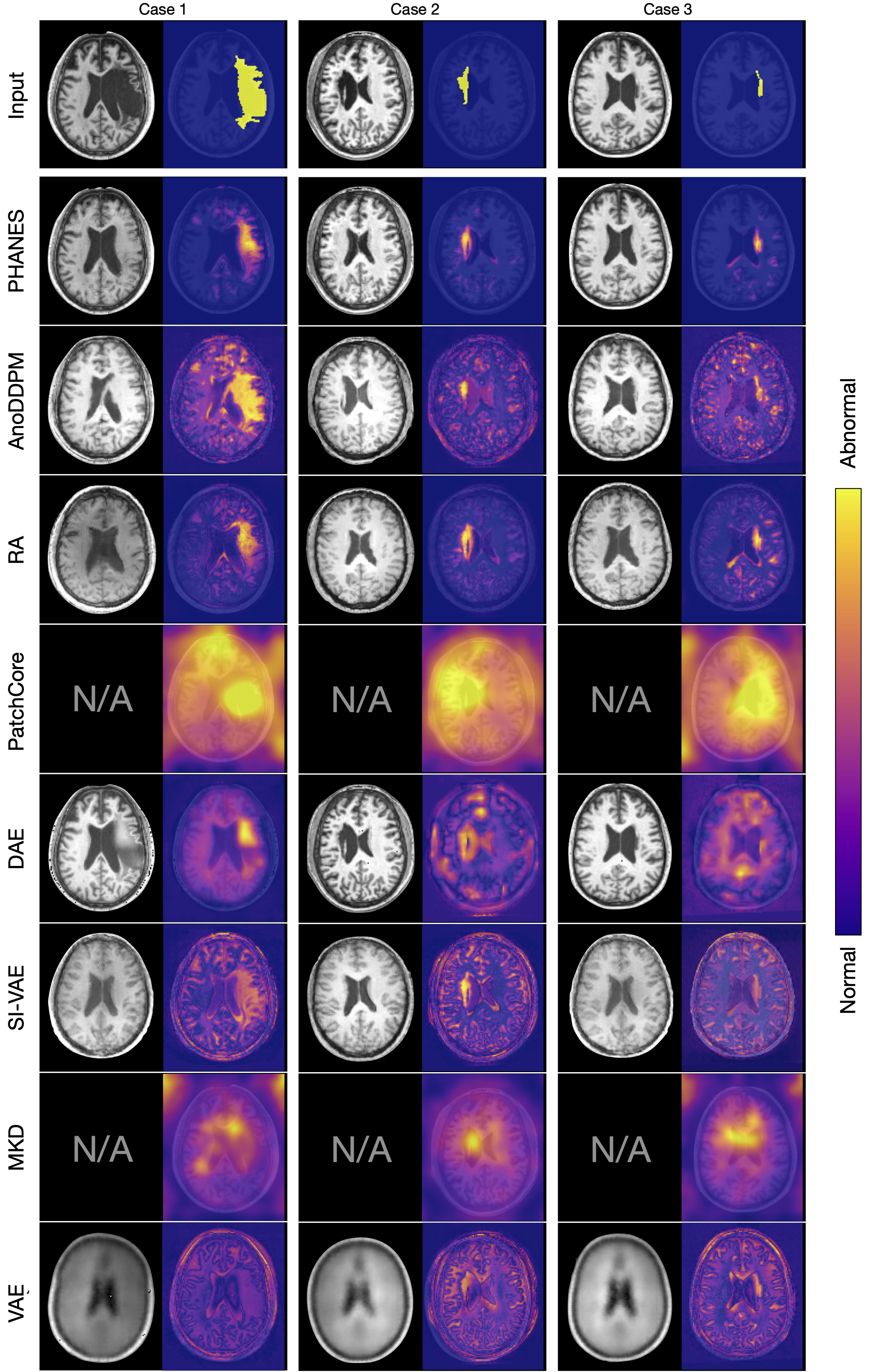} 
   
     \caption{Stroke lesion segmentation. Full qualitative comparison. The different columns show different cases. The first row shows input images and expert annotations masks. The following rows shows reconstructions (if available) and anomaly maps of \textit{PHANES} (ours) and different baselines: 
     AnnoDDPM~\protect\footnote{\url{https://github.com/Julian-Wyatt/AnoDDPM}}, 
     RA, PatchCore~\protect\footnote{\url{https://github.com/amazon-science/patchcore-inspection}}, DAE~\protect\footnote{\url{https://github.com/AntanasKascenas/DenoisingAE/}}, SI-VAE~\protect\footnote{\url{https://taldatech.github.io/soft-intro-vae-web}}, MKD~\protect\footnote{\url{https://github.com/rohban-lab/Knowledge_Distillation_AD}}, and VAE~\protect\footnote{\url{https://github.com/MIC-DKFZ/vae-anomaly-experiments/}}.\label{fig::quali_appendix}}
    \end{minipage}
\end{figure}   

%
\end{document}